# Monopole density around static color sources [*]

Harald Markum, Wolfgang Sakuler and Stefan Thurner

Institut für Kernphysik, Technische Universität Wien, A–1040 Vienna, Austria

We analyze the vacuum structure with respect to magnetic monopoles of quenched QCD in the presence of static color sources. Distributions of the monopole density around static quarks and mesons are computed in both phases of QCD. We observe a suppression of the monopole density in the vicinity of external sources. In the confinement phase the density of color magnetic monopoles is reduced along the flux tube between a static quark-antiquark pair.

**1.** A challenging problem in QCD is to identify its confinement mechanism and to understand how it works. It has been conjectured that the QCD vacuum is a coherent state of color magnetic monopoles [1]. In this picture color electric charges are confined due to the dual Meissner effect, where magnetic monopoles force the color electric field between a quark-antiquark pair into an Abrikosov flux tube, leading to the linear confinement potential.

In order to investigate monopole currents one has to project $SU(3)$ onto its abelian degrees of freedom such that an abelian $U(1) \times U(1)$ theory remains [2]. For this aim one has to perform an appropriate gauge fixing. However, the gauge fixing procedure is not unique. In our calculations we employ the so-called maximal abelian gauge which is the most favorable for our purposes. After gauge fixing quenched QCD can be regarded as a theory of color charges and color magnetic monopoles. In the present work we focus on distributions of monopole densities around static quarks and mesons which are obtained by computing correlation functions of Polyakov loops and local monopole densities in both phases of QCD.

**2.** A gauge transformation of a gauge field element $U(x,\mu)$ is given by

$$\tilde{U}(x,\mu) = g(x)U(x,\mu)g^{\dagger}(x+\hat{\mu}) ,\qquad(1)$$

where $g(x) \in SU(3)$. The maximal abelian gauge

[*]Supported in part by "Fonds zur Förderung der wissenschaftlichen Forschung" under Contract No. P9428-PHY.

is imposed by maximizing the functional

$$R = \sum_{x,\mu,i} |\tilde{U}_{ii}(x,\mu)|^2 . \qquad(2)$$

To extract abelian parallel transporters one has to perform the decomposition

$$\tilde{U}(x,\mu) = c(x,\mu)u(x,\mu) , \qquad(3)$$

with

$$\begin{aligned}
u(x,\mu) &= \text{diag}\,[u_1(x,\mu), u_2(x,\mu), u_3(x,\mu)] , \\
u_i(x,\mu) &= \exp\left[i \arg \tilde{U}_{ii}(x,\mu) - \frac{1}{3}i\phi(x,\mu)\right] , \\
\phi(x,\mu) &= \sum_i \arg \tilde{U}_{ii}(x,\mu)\Big|_{\text{mod}\,2\pi} \in (-\pi,\pi] .
\end{aligned} \qquad(4)$$

Since the maximal abelian subgroup $U(1) \times U(1)$ is compact, there exist topological excitations. These are color magnetic monopoles which have integer valued magnetic currents on the links of the dual lattice:

$$m_i(x,\mu) = \frac{1}{2\pi} \sum_{\square \ni \partial f(x+\hat{\mu},\mu)} \arg u_i(\square) , \qquad(5)$$

where $u_i(\square)$ denotes a product of abelian links $u_i(x,\mu)$ around a plaquette $\square$ and $f(x+\hat{\mu},\mu)$ is an elementary cube perpendicular to the $\mu$ direction with origin $x+\hat{\mu}$. The magnetic currents form closed loops on the dual lattice as a consequence of monopole current conservation. Finally the local monopole density is given by

$$\rho_i(x) = \frac{1}{4V_4} \sum_\mu |m_i(x,\mu)| . \qquad(6)$$

## Color Magnetic Monopole Density in a Static Meson

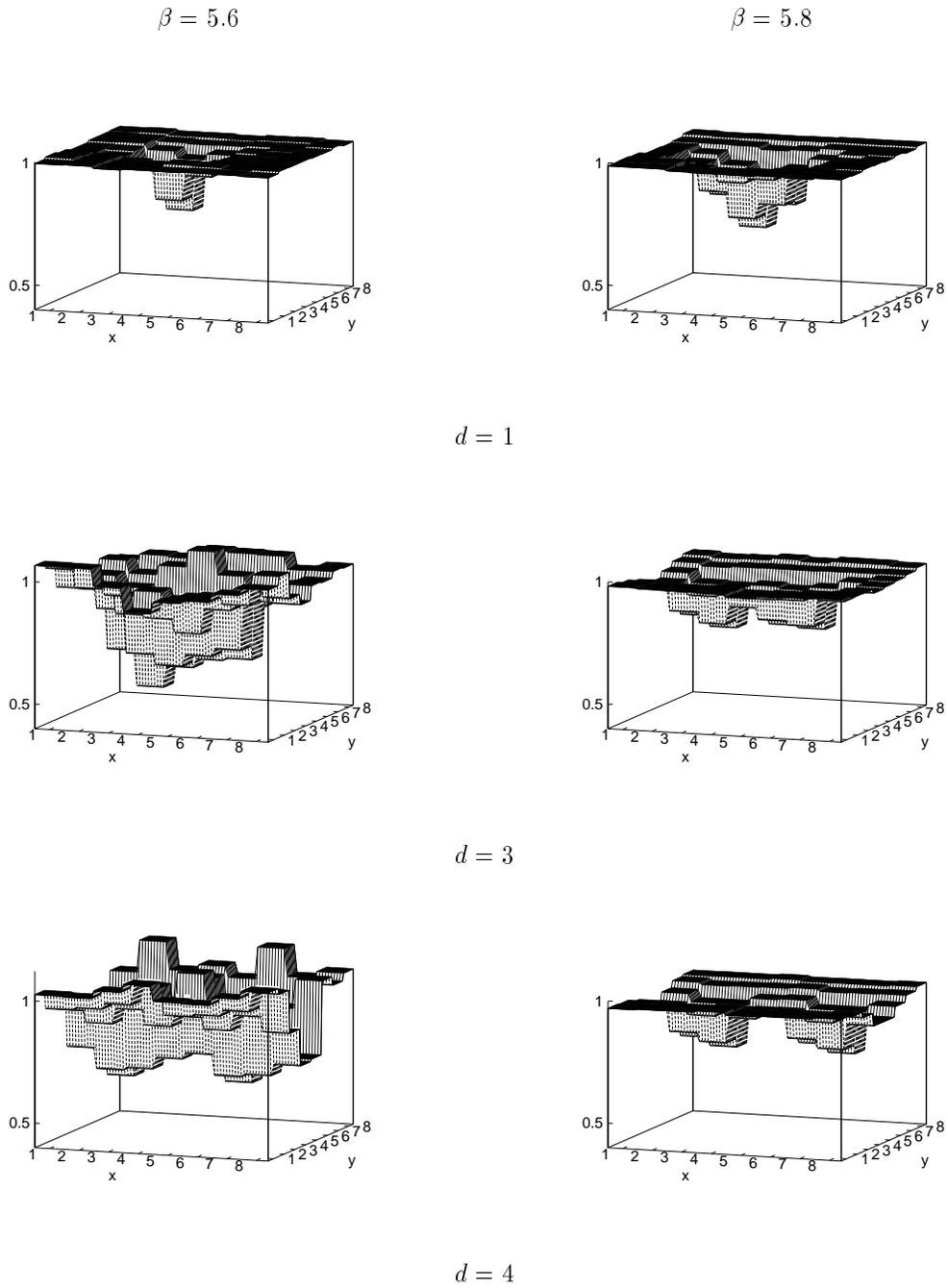

$d = 1$

$d = 3$

$d = 4$

Figure 2. Abelian monopole density around a static quark-antiquark pair for $q\bar{q}$-distances $d = 1, 3$ and 4. The suppression along the flux tube between the $q\bar{q}$-pair is in agreement with the expectation in the dual superconductor model.

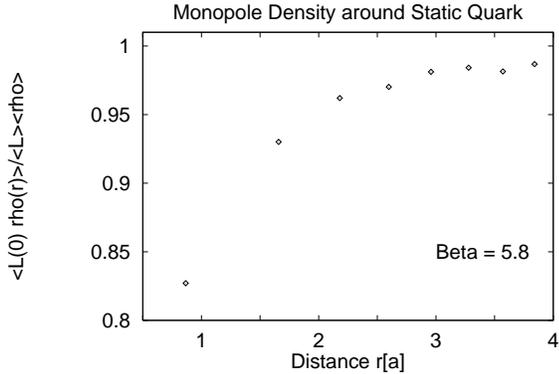

Figure 1. Color magnetic monopole density around a static quark at $\beta = 5.8$ (deconfinement phase).

A static quark is represented by the Polyakov loop $L(\vec{r})$ which describes the propagation of a charge with infinite mass. To measure the distribution of color magnetic monopoles around a static quark-antiquark pair and a single static quark, we calculate the correlation functions $\langle L(0)L^\dagger(d)\rho(r)\rangle$ and $\langle L(0)\rho(r)\rangle$, respectively, with $\rho(r) = \frac{1}{3}\sum_{i=1}^{3}\rho_i(r)$ and normalize them to their corresponding cluster values.

**3.** Our simulations were performed on an $8^3 \times 4$ lattice with periodic boundary conditions using the Metropolis algorithm. The observables were studied in pure QCD at inverse gluon coupling $\beta = 6/g^2 = 5.6$ (confinement) and 5.8 (deconfinement) employing the $SU(3)$ Wilson plaquette action for the gluons. We made 50000 iterations and measured our observables after every 50th iteration. Each of these 1000 configurations was subjected to 300 gauge fixing steps enforcing the maximal abelian gauge [3].

In Fig. 1 the density of color magnetic monopoles around a single static quark in the deconfinement phase is shown. A significant decrease in the vicinity of the quark is observed. In the confinement region the (unnormalized) correlation function should vanish since a single static quark is a forbidden state. This behavior was confirmed in our calculations.

The distributions of the color magnetic densities around static mesons for $q\bar{q}$-separations $d = 1, 3, 4$ are displayed in Fig. 2. A local suppression of the monopole densities around the charges is clearly visible in both phases. In the confinement phase the valley of reduced density reflects the formation of a flux tube between the static $q\bar{q}$-pair, whereas in the deconfinement phase the effect is only peaked at the position of the charges. Our work thus provides further evidence that monopoles dynamically rearrange the background color-electric field into an effective flux tube in the confinement phase of pure QCD. This behavior is in perfect agreement with the dual superconductor picture. There a diminished monopole density is expected inside the flux tube due to the restauration of the abelian symmetry which is spontaneously broken in the vacuum outside where monopoles are condensed.

**4.** Our results clearly show a change of the topological properties of the lattice vacuum across the deconfining phase transition with respect to monopoles. A qualitatively similar effect was observed in an investigation of instanton densities around static quarks [4]. Motivated by these results we plan to study correlations between monopoles and instantons to further clarify the role of topological excitations in the confinement mechanism.